\documentclass[12pt,preprint]{aastex}

\def\rjup{ R_{J}}

\def\rsun{\rm R_{\odot}}
\def\rstar{R_{\star}}
\def\mstar{M_{\star}}
\def\msun{\rm M_{\odot}}

\def\hst{\emph{HST}}

\begin{document}

\title{Precise Estimates of the Physical Parameters for the Exoplanet System HD~17156 Enabled by HST FGS Transit and Asteroseismic Observations}

\author{	Philip Nutzman\altaffilmark{1},
		Ronald L. Gilliland\altaffilmark{2},
		Peter R. McCullough\altaffilmark{2},
		David Charbonneau\altaffilmark{1},
		J{\o}rgen Christensen-Dalsgaard\altaffilmark{3},
		Hans Kjeldsen\altaffilmark{3},
		Edmund P. Nelan\altaffilmark{2},
		Timothy M. Brown\altaffilmark{4},
		Matthew J. Holman\altaffilmark{1}
		}

\altaffiltext{1}{Harvard-Smithsonian Center for Astrophysics, 60 Garden St., 
Cambridge, MA 02138}

\altaffiltext{2}{Space Telescope Science Institute, 3700 San Martin Drive, Baltimore, MD 21218, USA}

\altaffiltext{3}{Department of Physics and Astronomy, Aarhus University, DK-8000 Aarhus C, Denmark}

\altaffiltext{4}{Las Cumbres Observatory Global Telescope, Goleta, CA 93117}

\email{pnutzman@cfa.harvard.edu}

\begin{abstract}
We present observations of three distinct transits of HD 17156b obtained with the Fine Guidance Sensors (FGS) on board the \emph{Hubble Space Telescope} (\hst)\footnote[5]{Based on observations with the NASA/ESA Hubble Space Telescope obtained at the Space Telescope Science Institute, which is operated by the Association of Universities for Research in Astronomy, Incorporated, under
NASA contract NAS5-26555.}.  We analyzed both the transit photometry and previously published radial velocities to find the planet-star radius ratio $R_p/\rstar = 0.07454 \pm 0.00035$, inclination $i=86.49 ^{+0.24}_{-0.20}$ deg, and scaled semi-major axis $a/\rstar = 23.19 ^{+0.32}_{-0.27}$. This last value translates directly to a mean stellar density determination $\rho_\star = 0.522 ^{+0.021}_{-0.018} \rm ~g~cm^{-3}$.  Analysis of asteroseismology observations by the companion paper of Gilliland et al. (2009) provides a consistent but significantly refined measurement\footnote[6]{While this work was in press, the companion work revised this measurement to $\rho_\star = 0.5301 \pm 0.0044 \rm ~g~cm^{-3}$.  We note that the change, which is less than 0.2 of the error, would translate to a stellar radius change of less than 0.001 ${\rm R}_{\odot}$ and does not materially affect the results presented in this paper.} of $\rho_\star = 0.5308 \pm 0.0040 \rm ~g~cm^{-3}$.  We compare stellar isochrones to this density estimate and find $\mstar = 1.275 \pm 0.018 ~\msun$ and a stellar age of $3.37^{+0.20}_{-0.47}$ Gyr.  Using this estimate of $\mstar$ and incorporating the density constraint from asteroseismology, we model both the photometry and published radial velocities to estimate the planet radius
$R_p= 1.0870~\pm~0.0066~\rjup $ and the stellar radius $\rstar = 1.5007 \pm 0.0076 ~\rsun$.  The planet radius is larger than that found in previous studies and consistent with theoretical models of a solar-composition gas giant of the same mass and equilibrium temperature.  For the three transits, we determine the times of mid-transit to a precision of 6.2 s, 7.6 s, and 6.9 s, and the transit times for HD 17156 do not show any significant departures from a constant period. The joint analysis of transit photometry and asteroseismology presages similar studies that will be enabled by the NASA \emph{Kepler} Mission. 

\end{abstract}

\section{Introduction}

Much progress in the study of extrasolar planets has been driven by the
discovery and characterization of transiting planet systems.  Follow-up
observations of transiting systems have allowed for the measurement of
planetary transmission and emission features (Charbonneau et al. 2002, 2005;
Deming 2005;  Grillmair et al. 2007; Richardson et al. 2007; Swain et al.
2008; Tinetti et al. 2007; Vidal-Madjar et al. 2003), phase variations in
planetary brightness (Harrington 2006; Knutson et al. 2007), and constraints
on the projected spin-orbit alignment angle (Winn et al. 2005).  Proper
interpretation of the above results relies on accurate determinations of basic
planetary parameters such as the planet radius, inclination, and planet-star
radius ratio, obtained through high precision transit photometry.  In addition
to these parameters, precise transit photometry provides accurate transit time
measurements, which can be used to search for timing perturbations caused by
additional planetary companions (e.g, Holman et al. 2005; Agol et al. 2005).

The discovery that HD 17156b, originally identified via Doppler observations by Fischer et al. (2007), transits its host star (Barbieri et al. 2007) has brought an exceptional system into the sample of transiting planets; its period (P= 21.2 days) and eccentricity (e=0.68) are second largest (to HD~80606b; Naef et al. 2001, Moutou et al. 2009) among all currently known transiting planets. Its high orbital eccentricity is of considerable interest to modelers of planetary formation and migration.  In particular, planet-planet scattering scenarios (e.g., Rasio \& Ford 1996, Chatterjee et al. 2008) or the Kozai effect (e.g., Wu \& Murray 2003; Takeda \& Rasio 2005; Fabrycki \& Tremaine 2007) offer a possible explanation for the high eccentricity and also predict that the planetary orbital axis may not necessarily be well aligned relative to the stellar spin axis.   Perhaps surprisingly, Rossiter-McLaughlin studies of Cochran et al. (2008), Barbieri et al. (2009), and Narita et al. (2009) have found projected spin-orbit angles consistent with zero misalignment for the HD 17156 system. Previous photometric studies of HD 17156b (Irwin et al. 2008, Gillon et al. 2008, Barbieri et al. 2009) have determined the planetary radius to be below the theoretical expectation for a solar composition ball of the same mass and equilibrium temperature ($R_p = 1.1 ~\rjup$), which suggests that the planet may be significantly enriched in heavy-elements.  We note however that the discrepancy with $R_p= 1.1 ~\rjup$ in each study is $1 \sigma$ or less.  

Unfortunately, ground-based photometry of HD~17156 is susceptible to systematic errors due to the paucity of suitably bright, nearby comparison stars, which can confound the modeling of this system's relatively shallow transit signal ($0.5 \%$).  An opportunity remains for space-based photometry to significantly improve the determination of fundamental planetary parameters.  
In this paper, we report \emph{Hubble Space Telescope }(\hst) transit observations of HD 17156b obtained with the Fine Guidance Sensors (FGSs).  FGS science observations became common during 2008-2009 after failures of the STIS, ACS, and NICMOS instruments; these particular observations were scheduled as part of a major FGS program to explore asteroseismology of a transiting planet host star (see the companion paper of Gilliland et al. 2009b).  HD 17156 was selected for this program due to its brightness and its location in the continuous viewing zone of Hubble during the late 2008 to early 2009 period.  
The detection of stellar oscillations is very challenging due to the long observational window required for the proper frequency resolution (roughly $1 \mu$Hz) and due to the small amplitude photometric variations that one seeks to observe (less than 10 ppm).  However, the detection of several oscillation modes can yield constraints on the stellar density to better than $1\%$ and, in many cases, the stellar age to better than $10\%$ (see e.g. Brown \& Gilliland 1994).  

There are two principal goals for this study: (1) to present precise photometry of three new transit light curves and use these to significantly refine the planetary and stellar parameters for HD~17156; and  (2) to incorporate the asteroseismology constraint on the stellar density of Gilliland et al. (2009b) into the transit modeling and further improve the determination of system parameters.  This paper is organized as follows.  In \S 2 we discuss the reduction and processing of the FGS observations, while in \S 3 we describe the light curve modeling.  In \S 4 we use the results from the data analysis together with an analysis of stellar-evolutionary models to determine stellar and planetary properties.  In \S 5 we summarize and discuss our findings.

\section{Observations and Reduction}

\subsection{FGS Transit Photometry}
We obtained FGS data for three separate transits of HD~17156b (UT 2008 Nov 07, UT 2008 Dec 19, and UT 2009 Feb 21).
These observations complemented a ten day run of nearly continuous observations from 2008 Dec 21-31 (the asteroseismology run), as well as additional observations for the calibration of the detector deadtime and background flux. HST has three FGS instruments, two of which are used for pointing \hst, with a third that can be used for science observations.  Each sensor uses four photomultiplier tubes (PMTs) which can be used as high-cadence photometers (see Schultz et al. 2004, Bean et al. 2008 for previous examples of FGS transit observations).  Our observations employed the FGS2r instrument and the 440-710 nm \emph{F583W} filter.  The effective wavelength of the FGS detector + filter is 583 nm. (We also obtained high resolution observations of HD~17156 in FGS Transfer mode; see \S 4.3).
 
FGS records photon counts at a cadence of 40 Hz in each of the four PMTs.  The measured photon count rates must be corrected for detector deadtime, which results from the PMTs' inability to detect newly arriving photons during an interval following a previously detected photon.  For the high flux of HD 17156, the detector deadtime suppresses flux \emph{and} flux variations by more than $10\%$, thus making an accurate calibration of the deadtime correction absolutely critical for transit modeling.

To each of the 4 PMTs, we apply a deadtime correction of the form:
\begin{equation}
 C_{T,i}=C_{M,i}/(1.0 - C_{M,i} (T_{D,i}/T_I))
\end{equation}
where $C_{M,i}$ and $C_{T,i}$ ($i$=1,2,3,4) are, respectively, the measured and deadtime-corrected counts for the i$^{\mathrm{th}}$ PMT during the integration time $T_I$ = 0.025 seconds.  $T_{D,i}$ are the deadtime coefficients for each PMT.  These coefficients have been calibrated through a careful comparison of the relative count rates with FGS2r of two stars of similar spectral type, differing by 4.32 magnitudes, with excellent STIS spectrophotometry available, and magnitudes from slightly brighter than 
HD 17156 to much fainter (Gilliland et al. 2009a).  

Gilliland et al. (2009a) give an error budget of $-0.5 \%$ to $+0.9\%$ for the deadtime coefficients, which corresponds to an error range in flux ratios of $-0.32 \%$ to $+0.56 \%$.  For HD~17156b's transit depth of $0.55 \%$, this error budget translates to changes in transit depth of -18 to +31 ppm.  As we will discuss below, the transit photometry gives a $1\sigma$ error on the transit depth of 55 ppm, as determined from a Markov Chain Monte Carlo (MCMC) analysis (which neglects the uncertainty in the deadtime correction).  Therefore, we conclude that systematics associated with an uncertain deadtime correction do not significantly impact the results of our analysis. For the analysis below, we sum over the four PMTs' (deadtime corrected) flux measurements and then sum into 30-second bins.  We found no advantage to analyzing the data at a higher time resolution, or for select sub-sums of the 4 PMTs.

Each of the three transit observations span 8-9 hours, with at least two hours of observations before and after the 3 hour transit.  As HD 17156 was not in \hst's continuous visibility zone (CVZ) for these observations, a portion of each 96-minute \hst~ orbital period is interrupted.  For the Nov 7, 2008 and Feb 21, 2009 transits, we discarded the first orbit of data as these showed anomalous photometric offsets and variability not present in the other orbits.  For the Nov 7, 2008 transit, we also discarded the first half of the second orbit, during which time HST passed through the South Atlantic Anomaly, resulting in a positive bump in the flux measurements due to charged particle events.  In total, our observations consist of $6 + 6 +6 = 18$ usable HST orbits of data with an average duty cycle of roughly $2/3$.  

To estimate changes in the background flux, we used the flux measurements from FGS3 which was trained on its guide star that is roughly 150 times fainter than HD 17156.  We assumed that changes in the count rate for FGS3 were due to changes in background count rate and used this as a proxy for changes in the FGS2r background count rate.  We found that correcting for this change in count rate had negligible effect on the photometry, except that it removed an anomalous ``bump'' in brightness (of relative amplitude $ 2 \times 10^{-4}$) in the raw photometry of the 4th orbit of the Dec 19 observations.

\subsection{\hst~Orbital Flux Variation and Correction}

Each observational sequence within a single HST orbit begins with the two guide FGSs turning on and establishing fine lock on guide stars. The High Voltage (HV) on FGS2r is then turned on, after which HD 17156 flux measurements increase rapidly to $97 \%$ of the full count rate (within 0.1 s of turn on).  Over the following 5 minutes, the counts gradually ramp up to the full count rate.  For uniformity in our data treatment, we choose the first 30 second flux sum of each orbit to begin exactly 21.5 seconds following HV turn on.  We describe our handling of the HV ramp later in this section. 

The most important systematic effects in the flux data are prominent periodic variations at the 96 minute HST orbital period.  The ``orbital waveform'' is a repeating signal with semi-amplitude $0.1 \%$ that evolves only modestly over the course of each 8-9 hour transit observation.  Over much longer timescales, however, the waveform evolves significantly (see also Gilliland et al. 2009b), so that the correction for this systematic must be handled individually for each of the three transit observations.

To correct for the orbital waveform, we fit the out-of-transit (OOT) data with a polynomial function of the HST orbital phase.  In an ideal treatment, the orbital waveform correction parameters would be simultaneously modeled with the transit parameters, to address directly the impact of uncertainties in the correction on the transit parameters.  However, we found this approach to be computationally impractical due to the tens of extra parameters that the orbital waveform correction introduces to the modeling.  The approach we adopted was to fix the waveform correction to that determined from the OOT data.  The phase was determined modulo a fixed orbital period ($P_{HST} = 95.9184$ minutes).  Although the HST orbital period decays at a significant rate (more than 20 seconds from Nov 7 to Feb 21), we found that using a fixed period sufficed for our purposes.  We experimented with taking into account HST orbital decay, but found that this led to negligible differences in the data processing compared to using the fixed HST period period given above.
The orbital waveform corrections were determined separately for each of the three HST visits via the following iterative process.  For each given visit, we determined an initial polynomial fit to the OOT data.  This initial fit was used to determine the flux offsets for each of the HST orbits.  These offsets were divided from the data and a new polynomial fit was determined.  We iterated this process 3 times.

To estimate the ``optimal'' degree for the polynomial fitting function, we employed a cross-validation test (see e.g. Mandel et al. 2009, who apply cross-validation to SN Ia light-curve inference).  The purpose of cross-validation is to avoid over-fitting data by including an unjustified number of model parameters.  Cross-validation is performed by fitting a model to a subset of the data (called the ``training set''), and assessing the prediction error of this model on the remaining portion of the data.  To improve the test, it is common to perform multiple rounds of cross-validation, using different subsets of the data as the training set.    We divided the OOT data into the 3 out-of-transit HST orbits, and performed three rounds of cross-validation.  For each round of cross-validation, one orbit was used as the training set to derive the polynomial fitting function, and the prediction error of this polynomial model was determined for the remaining two orbits.  We assessed the root-mean-square prediction error for a range of polynomial degrees, and found the degree for which the rms prediction error was minimized.  The optimal degrees were determined to be 9, 8, and 8 for the Nov 7, Dec 19, and Feb 21 transits, respectively.  Note that once the optimal degree was determined, we then fit the polynomial of this degree to all OOT data (per visit).

The ramp up in sensitivity during the 5 minutes following HV ramp-up is another important systematic.  However, with the exception of the first orbit of the Dec 19 transit observations, all HV ramp-ups within a set of transit observations occur at the same HST orbital phase.  The result is that any attempt to separately handle the HV ramp-up and orbital correction would suffer from the degeneracy in orbital phase.  Our approach is to discard the first 5 minutes of the first orbit of the Dec 19 transit, and to allow our polynomial function of phase to correct for both the HV ramp-up and orbital variations.

All data analyzed below has been divided by the best-fit orbital waveform determined via the process described above.

\begin{figure} 
\epsscale{1.0}
\plotone{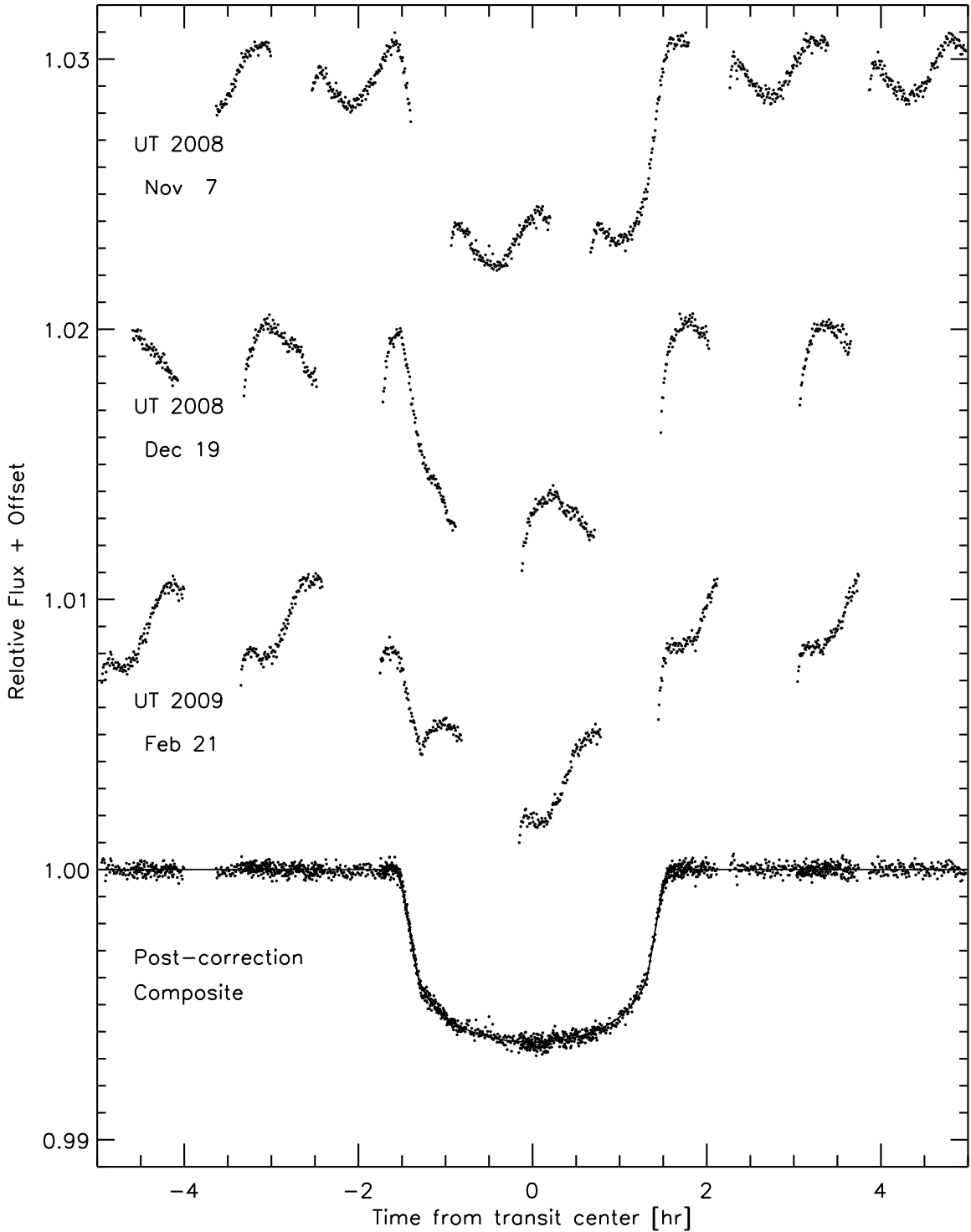}
\caption{FGS photometry of three HD 17156b transits.  \emph{Top}: Raw photometry, offset for clarity of display.  \emph{Bottom}: Folded and corrected photometry.  The data have been background subtracted, corrected for HST orbital variations, and divided by a quadratic function of time as described in \S 2 and \S3. } 
\end{figure}
\clearpage


\section{Joint Radial Velocity and Light Curve Analysis}

In this section we describe a joint analysis of published radial velocity data and the new FGS transit light curves.  We include the radial velocity data published by Fischer et al. (2007), and Winn et al. (2009), excluding any data affected by the Rossiter-Mclaughlin effect.  The Fischer et al. (2007) data consist of nine velocities obtained with the High Dispersion Spectrograph on the Subaru 8 m telescope and 24 velocities obtained with the HIRES spectrograph on the Keck I 10 m telescope.  Winn et al. (2009) adds 10 new Keck/HIRES velocities and re-measures all 34 Keck/HIRES velocities using a refined reduction procedure, which includes the use of a new HD~17156 spectral template.  Our analysis makes use of all the Subaru data and the re-reduced Keck/HIRES velocities and errors reported by Winn et al. (2009). 

The joint model for the radial velocity data and photometry consists of 21 free parameters: 5 describing the planetary orbit ($P$, $e$, $\omega$, $K$, $T_{peri}$), 5 describing the transit light curve ($a/\rstar$, $R_p/\rstar$, $i$, and two quadratic limb darkening coefficients), two radial velocity zero-point offsets, and 9 parameters describing corrections to the photometry.  To produce quadratically limb-darkened transit light curves, we employed the analytic formulas of Mandel \& Agol (2002).  The 9 photometric correction parameters consist of coefficients to a parabolic function of time for each set of transit observations (one parabola, three coefficients for each HST visit).  Note that this photometric correction is modeled simultaneously with the planetary orbit and transit parameters, in contrast to the calibration steps discussed in the previous section, which were determined solely from out-of-transit data.  This correction proved necessary because long-term residual trends are apparent in each set of transit observations after dividing out the HST orbital waveform correction and dividing out the transit fit determined from the combination of the 3 transit light curves.  We note that the impact of this correction is modest; peak-to-peak the magnitude of the corrections are $0.05 \%$ and the correction coefficients have only small correlation with the transit and planetary orbit parameters.   

For parameter estimation, we employ a Metropolis-style Markov Chain Monte Carlo algorithm (MCMC; see, e.g., Ford 2005, Holman et al. 2006 and references therein).  Our acceptance probability for newly proposed links in the Markov chain is based on the $\chi^2$ statistic:

\begin{equation}
\chi^2 = \frac{(\mathrm{f}_{\rm obs,i} - \mathrm{f}_{\rm mod,i})^2}{\sigma_i^2},
\end{equation}
where $\mathrm{f}_{\rm obs,i}$ is the $i^{th}$ FGS flux measurement, $\mathrm{f}_{\rm mod,i}$ is the $i^{th}$ model flux, and $\sigma_i$ is the $i^{th}$ measurement error.  $\sigma_i$ was scaled such that the $\chi^2$ per degree of freedom equals 1.



Median values and the central $68.3 \%$ confidence limits for the transit parameters, radial velocity parameters, and other directly observable quantities are reported in Table 1.  These quantities are affixed with the label ``A" to emphasize that they are derived independently from external assumptions on stellar properties.  We discuss these results and compare to previously published determinations in \S 5.

\section{Stellar and Planetary Properties}

\subsection{Stellar Isochrone Modeling}

High quality transit light curves provide an important constraint on the stellar mean density\footnote{We note that this density constraint, while very useful, is less than perfect due to complicating factors such as star spots and plages, uncertain limb darkening coefficients, and uncertain orbital eccentricity.}, $\rho_\star$, through the transit parameter $a/\rstar$ and the application of Newton's version of Kepler's Third Law (see e.g. Seager \& Mall\`{e}n-Ornelas 2003).  When analyzing stellar evolution models, this determination of $\rho_\star$ is a valuable complement to the spectroscopically determined properties $T_{\rm eff}$ and [Fe/H] (see for example Sozzetti et al. 2007; Torres et al. 2008).  In this section, we describe our determination of stellar properties via a comparison of these observables to stellar isochrones. 

We consulted the Yonsei-Yale (${\rm Y}^2$) stellar evolution models by Yi et al. (2001) and Demarque et al. (2004).  These evolution models use the OPAL equation-of-state and opacities and incorporate the diffusion and settling of helium.  Overshoot is included from convective cores over a distance that increases with mass in a step-wise fashion to 0.2 pressure scale heights.
The abundances $X$ and $Z$ by mass of hydrogen and heavy elements
are assumed to be related by $X = 0.77 - 3Z$; the composition is related to [Fe/H], taking the present
solar surface composition to satisfy $Z/X = 0.0253$.  

As observational inputs, we adopted $T_{\rm eff} = 6079 \pm 80$ K, [Fe/H] = $ +0.24 \pm 0.05$, and the absolute magnitude $M_V$ = $3.80 \pm 0.12$. 
The above determinations of [Fe/H] and $T_{\rm eff}$ are from Fischer et al. (2007), but with increased error bars, following Winn et al. (2009).  We also note agreement between these values of [Fe/H] and $T_{\rm eff}$ and those found by Ammler-Von Eiff et al. (2009). We converted the light curve constraint on $a/\rstar$ (see section \S 3 and Table 1) to the corresponding constraint on stellar mean density, $\rho_{\star}= 0. 524^{+0.021}_{-0.018}$.

 We computed isochrones over a 2D grid of metallicity and stellar age, with metallicity ranging from [Fe/H]=0.19 to 0.29 in steps of 0.005 dex and age ranging from 1 to 5 Gyr in steps of 0.01 Gyr.  Each outputted model was weighted in proportion to exp$(-\Delta \chi^2_{\star}/2)$ with
\begin{equation}
\Delta \chi^2_{\star} =
\left[ \frac{\Delta{\rm [Fe/H]}}{\sigma_{{\rm [Fe/H]}}} \right]^2 +
\left[ \frac{\Delta T_{\rm eff}}{\sigma_{T_{\rm eff}}} \right]^2 +
\left[ \frac{\Delta \rho_\star}{\sigma_{\rho_\star}} \right]^2,
\end{equation}
where the $\Delta$ values represent deviations from the observed and model calculated values.  Following Winn et al. (2009), we handled the asymmetric error in $\rho_{\star}$ by using different values of $\sigma_{\rho_{\star}}$ depending on the sign of the deviation.  The weight was then multiplied by a factor to take into account the number density of stars along each isochrone, assuming a Salpeter mass function. This analysis yielded $\mstar = 1.275 \pm +0.018~ \msun$, and a stellar age of $3.38^{+0.20}_{-0.47}$ Gyr.

Using the above constraint on stellar mass and employing Newton's version of Kepler's Third Law, we determine the semimajor axis, $a$, allowing us to translate the dimensionless parameter from our MCMC analysis, $a/R_\star$ into physical units.  This final step yields $R_\star=1.508 \pm 0.021~\rsun$ and $R_p = 1.095 \pm 0.020~\rjup$, where the reported errors take into account the uncertainty in stellar mass.  The value for these and other parameters are reported in Table 1 and discussed in \S 5.

\subsection{Incorporating Information from $\rho_\star$ as Determined by Asteroseismology}

The companion paper of Gilliland et al. (2009b) presents a robust detection of p-mode oscillations in HD~17156, the first such measurement for a transiting planet host star.  Because these asteroseismology measurements facilitate the measurement of $\rho_\star$, HD~17156 is the first star for which a direct determination of $\rho_\star$ was obtained using both asteroseismology and transit photometry.  Gilliland et al. (2009b) report the determination of $\rho_\star = 0.5308 \pm 0.0040 $g cm$^{-3}$, based on analysis of identified frequencies for p-modes of degree $l$ = 0, 1, and 2.  This estimate is 4 times more accurate than our transit photometry determination of $0.522^{+0.021}_{-0.018}$ g cm$^{-3}$, though we note that the determinations are mutually consistent.

The asteroseismology observations determine $\rho_\star$ and hence $a/\rstar$ ($ a/\rstar = 23.287 \pm 0.058)$ significantly more precisely than does the transit photometry.  When $a/\rstar$ is accurately constrained, the orbital inclination can be directly determined by the transit duration.  This is a significant benefit over the typical transit modeling scenario, which relies on a delicate measurement of the the ingress/egress shape to disentangle $a/\rstar$ from the inclination. Thus an accurate and independent constraint on $a/\rstar$ can significantly refine the determination of the inclination and hence the impact parameter of the transit chord.  Importantly, a precisely determined transit chord better informs the determination of the parameter $R_p/\rstar$, which is usually strongly covariant with $a/\rstar$.  These considerations motivate incorporating the asteroseismology determination of $a/\rstar$ as a Bayesian prior on the transit analysis.  

We, therefore, repeat the MCMC analysis of \S 3, but modify the $\chi^2$ statistic of equation (2) with an added term $(a/\rstar - 23.287)^2/0.058^2$.  The effect of this added constraint is to dramatically improve the estimates of the transit parameters.  The results from this investigation are reported in Table 2 and discussed in \S 5.  Note that the uncertainty for $a/\rstar$ in Table 2 is slightly smaller than the prior uncertainty; this reflects the relatively modest information content provided by the transit photometry for $a/\rstar$.

Though our transit analysis yielded a determination of the stellar mean density that is consistent with the asteroseismology determination, it is worth considering hypothetical scenarios which may lead to a discrepancy between density determinations for transiting planet hosts with asteroseismology observations.  Ford et al. (2008) have pointed out that an eccentric planet, if assumed to be on a circular orbit, would lead to a discrepancy between the transit-inferred and intrinsic $\rho_\star$.  Ford et al. suggest that this effect could be used to characterize the orbital eccentricities of planets discovered by the NASA \emph{Kepler} Mission (Borucki et al. 2007).  Below, we note three further scenarios that could lead to a discrepant transit determination of $\rho_\star$, and which are degenerate with the effect of unmodeled orbital eccentricity.  

The well-known expression for $a/\rstar$ in terms of the stellar density (see e.g., Seager \& Mall\`{e}n-Ornelas 2003) includes a generally neglected term that depends on the planet-star mass ratio $q = M_p/M_\star$.  Planet-mass objects will not yield a measureable discrepancy between transit and asteroseismology determinations of $\rho_\star$, but consistency between the determinations can provide a weak upper limit on $q$. We rearrange the expression for $a/\rstar$, solving for $q$: 
\begin{equation}
q = \left(\frac{a}{\rstar}\right)^3 \frac{3 \pi}{G P^2} \frac{1}{\rho_\star} - 1 
\end{equation}
Plugging in $a/\rstar$ as determined by transit photometry and $\rho_\star$ as determined by asteroseismology, we find $q= -0.008 \pm 0.042$, with the precision limited by the uncertainty in the photometrically determined $a/\rstar$.  For HD~17156b, which has a radial velocity-determined mass, this constraint is superfluous.  We note, however, that the \emph{Kepler} Mission is expected to provide both asteroseismology and transit light curves for many systems which will initially lack radial velocity data.  The above constraint could be useful as an initial diagnostic for the presence of a Jupiter-sized late M-dwarf masquerading as a transiting planet\footnote{On the other hand, objects showing a significantly non-zero $q$ may also be bright enough to present secondary eclipses detectable by \emph{Kepler}.}.  

Contamination of a transit light curve by third light will also lead to a discrepancy between the transit-determined and true $a/\rstar$.  Third light dilutes the transit depth, $\delta$, thus impacting the measurement of $a/\rstar$.  To first order, $a/\rstar$ depends most strongly on the transit duration, ingress/egress duration, and orbital period, however there is a modest dependence on transit depth  ($\propto \delta^{-1/4}$; see e.g., Carter et al. 2009).  To exploit this effect and thus constrain the amount of third light, we repeated the MCMC analysis of this section (i.e. with the Bayesian prior on $a/\rstar$), but included an additional free parameter $F_3$, the fraction of total flux contributed by a non-variable third object.  We defined $F_3$ such that the sum of fluxes for all three objects (HD~17156, HD~17156b, hypothetical third object) is unity.  This analysis yields $F_3 < 0.16$ at $95 \%$ confidence.  The following section also describes high angular resolution observations which were able to place stronger limits on the presence of stellar companions.

Finally, a common false-positive that plagues transit searches is caused by an eclipsing binary blended with a far brighter but unresolved star.  In this case, there is no relation between the transit and asteroseismology determinations of $\rho_\star$.  On the other hand, for planet candidates that show coincidence between the transit and asteroseismology determined $\rho_\star$'s, one has strong evidence against this false-positive scenario.  This tool could prove useful for vetting \emph{Kepler} candidates that are bright enough to obtain successful asteroseismology observations, before obtaining radial velocity follow-up.     

\subsection{High Angular Resolution Check for Stellar Companions}

We obtained high angular resolution observations of HD 17156 in FGS Transfer mode. In Transfer mode the FGS samples an object's interference fringes with 1 mas steps in both the X and Y channels (for details of the FGS operation see Nelan et al., 2010). The resultant fringe morphology and amplitude are compared to fringes obtained from observations of a point source calibration star of similar B-V color. We compared the HD 17156 fringes to those obtained from observations of UPGREN-69 (HIP 3354) with FGS2r. HD 17156 showed no departure from the UPGREN-69 fringes, i.e., HD 17156 appears to be a points source down to the angular resolution of FGS2r in both the X and Y channels.

The angular resoluton limits of FGS2r can be estimated from simulations of model binary systems. We estimate that FGS2r can begin to detect binary systems with projected separations of 8 to 10  mas for equal brightness components. Components with $\Delta V < 2$ can be detected for separations greater than 14 to 15 mas, while binaries composed of stars with $\Delta V \approx$ 3 need to have separations greater than about 20 mas for detection with FGS2r. We note that at the Hipparcos distance of 78 pc, 20 mas corresponds to 1.6 AU. The FGS non-detection of binarity in HD 17156, combined with the radial velocity data which precludes a stellar mass body with an orbital period $P < 2$ year, excludes any companion with $\Delta V < 3$.  Our results complement those of Daemgen et al. (2009) who found no companions with a projected separation greater than a few arc-seconds and down to $\Delta i' < 8$.


\subsection{Refined Ephemeris and Search for Transit Timing Variations}

To determine transit timings we performed a new analysis that models only the transit photometry.  We fix $e$ and $\omega$ at the best fit values determined from the joint radial velocity and transit photometry analysis.  The model for this analysis consists of the 5 transit light parameters and 9 photometric correction parameters described in \S 3.  Otherwise, the MCMC implementation is unchanged from \S 3.

The three precise transit timings (reported in Table 1) allow for a significant refinement in the transit ephemeris.  Our analysis includes 5 previously published transit timings which are tabulated in Winn et al. (2009).  We fit the transit timings to a linear ephemeris: $T_c[E] = T_0 + E \times P$.  We determined 
\begin{equation}
T_0 =2454884.028170  \pm 0.000073 ~{\rm [HJD]},
\end{equation}
\begin{equation}
P=21.2163979 \pm 0.0000159 ~{\rm days}.
\end{equation}

The residuals to this ephemeris are plotted in Figure 2.  We find no obvious deviations from a constant period.

\begin{figure} 
\epsscale{1.0}
\plotone{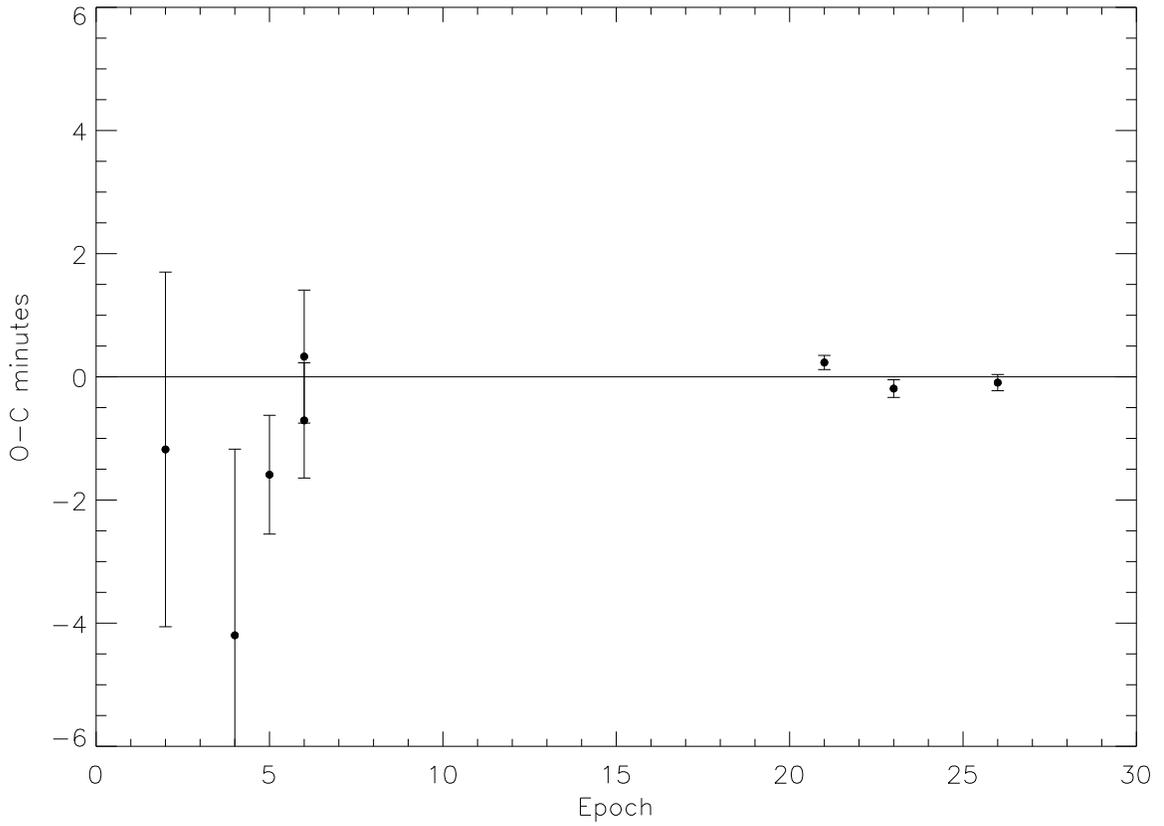}
\caption{Transit timing residuals for HD 17156b.  The calculated transit times, using the ephemeris given in \S 4.2, are subtracted from the observed times.} 
\end{figure}
\clearpage

\begin{deluxetable}{lccc}
\tabletypesize{\scriptsize}
\tablecaption{System Parameters of HD 17156\label{tbl:params}}
\tablewidth{0pt}

\tablehead{
\colhead{Parameter} & \colhead{Value} & \colhead{68.3\% Conf.~Limits} & \colhead{Notes}
}

\startdata
{\it Stellar parameters:} & & & \\
Mass, $\mstar$~[M$_{\odot}$]                    &  1.275 & $\pm 0.018$   &  B \\
Radius, $R_\star$~[R$_{\odot}$]   		&  1.508 & $\pm 0.021$ & C\\
Mean density, $\rho_\star$~[g~cm$^{-3}$]        &   0.522 &-0.018, +0.021   & A  \\
Age [Gyr] 					& 3.38 & -0.47, +0.20	 & B \\
& & & \\
{\it Transit ephemeris:} & & & \\
Reference epoch~[HJD]                           & 2454884.028170&  $\pm 0.000073$    & A  \\
Orbital period~[days]                           &  21.2163979     &  $\pm 0.0000159$  & A  \\
& & & \\

{\it Orbital parameters:} & & & \\
Velocity semi-amplitude, $K$~[m s$^{-1}$]	&  274.2	& $\pm 2.0 $ & A \\
$e$						&  0.6768	& $\pm 0.0034 $ & A \\
$\omega$~[deg]					&  121.71	& $\pm 0.43 $ & A\\
& & & \\

{\it Transit parameters:} & & & \\
Midtransit time on 2008~Nov~07~[HJD]            & 2454777.946341  &  $\pm  0.000081$    & A   \\
Midtransit time on 2008~Dec~19~[HJD]            & 2454820.378843   & $\pm  0.000108$ &   A\\
Midtransit time on 2009~Feb~21~[HJD]		& 2454884.028105 & $\pm  0.000103$  & A\\
Planet-to-star radius ratio, $R_p/R_\star$             & 0.07454   & 0.00035  & A  \\
Orbital inclination, $i$~[deg]                        &  86.49  &  -0.20, +0.24  &  A \\
Scaled semimajor axis, $a/R_\star$                     &  23.19   &  -0.27, +0.32& A  \\
Transit impact parameter, $b_{\rm I} \equiv \frac{1-e^2}{1 + e\sin{\omega}} a\cos i/R_\star$ &  0.477   & -0.029, + 0.023  &  A,D \\
Transit duration~[hr]                                 &  3.1435    & $\pm 0.0059$ &  A \\
Transit ingress or egress duration~[hr]               &   0.2717   & $\pm 0.0092$ &  A \\
Impact par. at superior conjunction, $b_{\rm II} \equiv \frac{1-e^2}{1 - e\sin{\omega}} a\cos i/R_\star$  & 1.774 & -0.107, +0.085 & A,D\\
& & &\\

{\it Planetary parameters:} & & & \\
$R_p$~[R$_{\rm Jup}$]                                 &  1.095   & $\pm 0.020$  &  C\\
$M_p$~[M$_{\rm Jup}$]				      &	 3.191	 & $\pm 0.033$	&  C\\
Surface gravity, $g$~[m s$^{-1}$]		      &  67.0	 & $\pm 2.4 $	&  A\\

\enddata

\tablecomments{(A) Determined from the joint analysis of transit photometry and radial velocity data. (B) Determined from isochrone analysis and including the stellar density constraint from the transit photometry.  The confidence limits reflect only the formal uncertainty in parameters and do not account for systematic errors or theoretical uncertainty in the stellar isochrones. (C) Determined from the light curve analysis, and assuming $M_\star = 1.275 \pm 0.018 ~$M$_{\odot}$.  R$_{\rm Jup} = 7.1492\times 10^9$ cm.  (D) As defined, the impact parameters at inferior and superior conjunction, $b_{\rm I}$ and $b_{\rm II}$, only yield an approximation of the minimum projected star-planet separation. However, for HD 17156b, the deviation from true minimum separation is negligible.}  

\end{deluxetable}

\begin{deluxetable}{lccc}
\tabletypesize{\scriptsize}
\tablecaption{Refined Parameters Using Density Constraint from Asteroseismology}
\tablewidth{0pt}

\tablehead{
\colhead{Parameter} & \colhead{Value} & \colhead{68.3\% Conf.~Limits} 
}

\startdata
{\it Stellar parameters:} & & & \\

Planet-to-star radius ratio, $R_p/R_\star$             & 0.07444   & $\pm 0.00022$    \\
Orbital inclination, $i$~[deg]                         &  86.573  &  $\pm 0.060$  \\
Scaled semimajor axis, $a/R_\star$                     &  23.281   &  $\pm 0.057$  \\
Transit impact parameter, $b_{\rm I} \equiv \frac{1-e^2}{1 + e\sin{\omega}} a\cos i/R_\star$ &  0.455   & $\pm 0.014$   \\
Impact par. at superior conjunction, $b_{\rm II} \equiv \frac{1-e^2}{1 - e\sin{\omega}} a\cos i/R_\star$  & 1.666 & $\pm 0.037$ \\
$R_p$~[R$_{\rm Jup}$]                                 &  1.0870   & $\pm 0.0066$  \\
$R_\star$~[R$_{\odot}$]		 		      & 1.5007 & $\pm 0.0076$  \\

\enddata

\tablecomments{Values are determined from an MCMC analysis which includes a prior on $a/\rstar$ corresponding to the mean stellar density determined from asteroseismology of HD~17156.  See \S 4.2}  

\end{deluxetable}

\section{Discussion}

We have presented \emph{Hubble} FGS photometry of three new transits of HD 17156b.  The high quality photometry allows us to revamp the characterization of the HD 17156 system.  In particular, we have measured the stellar radius directly from the transit light curves (with respect to a stellar mass of $1.27 ~\msun \pm 0.018$, as derived from the isochrone analysis), in contrast to most previous HD 17156 studies, which either resort to or cannot significantly improve on external determinations of the stellar radius.  Our stellar radius measurement, $R_\star = 1.508 \pm 0.021~\rsun$, is larger than but consistent with the previous determinations of Barbieri et al. 2009 ($ 1.44 \pm 0.08 ~\rsun$) and Winn et al. 2009 ($1.446^{+0.099}_{-0.067} ~\rsun$), with a factor of 4 improvement in precision relative to these previous studies.  When incorporating the asteroseismology constraint on the stellar density into the transit analysis, the radius determination is dramatically refined to $R_\star = 1.5007 \pm 0.0076$.   Our stellar radius determinations are consistent with a determination of $\rstar = 1.45 \pm 0.07 \rsun$ obtained using the Kervella et al. (2004) color-angular diameter relations and the observed parallax (see Barbieri et al. 2009).  The larger radius finding results, in part, from a larger stellar mass than that found in previous studies (1.24 $\msun$ in Barbieri et al. 2009; 1.263 $\msun$ in Winn et al. 2009).   

The larger stellar radius measurement leads, in turn, to a larger planetary radius measurement.  Our result of $R_p = 1.095 \pm 0.020~\rjup$ is larger than but consistent with the value $R_p= 1.02 \pm 0.08 ~\rjup$ found by Barbieri et al. 2009, and larger than but consistent the value $R_p= 1.023 ^{+0.070}_{-0.055} ~\rjup$ found by Winn et al. 2009.  When including the asteroseismology constraint, the planet estimate is further improved to 1.0870 $\pm 0.0066 \rjup$.  Our planet-star radius ratio measurement, $R_p/R_\star$= 0.07454 $\pm 0.00037$ (0.07444 $\pm 0.00022$ post asteroseismology constraint), is consistent with the earlier findings of Barbieri et al. 2009 and Winn et al. 2009, and thus the enlarged planetary radius can be fully attributed to the enlarged stellar radius. 

Compared to the inclination value of Barbieri et al. (2009), $i = 87.9 \pm 0.1$ deg, we find a significantly lower value of $i=86.49^{+0.24}_{-0.20}$ deg (86.573 $\pm 0.060$ deg postasteroseismology constraint).  This value is consistent with but much more precise than the Winn et al. (2009) value of $86.2^{+2.1}_{-0.8}$ deg.  Our precise inclination determination also enables us to investigate the possibility of secondary eclipse.  We determined the a posteriori distribution of 
\begin{equation}
b_{\rm II} \equiv \frac{1-e^2}{1 - e\sin{\omega}} a\cos i/R_\star
\end{equation}
which, for the HD 17156b system, gives an excellent approximation of the minimum projected star-planet separation at superior conjunction.  We find $b_{\rm II} = 1.774^{+0.085}_{-0.107}$ (1.666 $\pm 0.037$ post asteroseismology constraint) which is roughly $7 \sigma$ ($16 \sigma$) greater than $1 +R_p/R_\star$, and thus rules out the possibility of secondary eclipse.  Even though the existence of secondary eclipse is strongly disfavored, we conducted a search for evidence of secondary eclipse using data from the ten day asteroseismology run, which covered the expected phase of secondary eclipse.  The FGS passband is far too blue to be sensitive to thermal emission, while the reflected light signal, for an albedo of 1, is predicted to be less than 5 ppm.  Unfortunately, the FGS observations exhibit significant systematic variations of magnitude 100 ppm on timescales comparable to the expected eclipse duration ($\approx10$ hours).  During the expected phase of secondary eclipse, the data only excludes eclipse depths greater $>$ 150 ppm, which is far greater than the predicted signal. 

Previous studies have suggested that HD 17156b may be enriched in heavy elements (Irwin et al. 2008; Winn et al. 2009).  The models of Fortney et al. (2007) for a solar-composition planet of the mass of HD 17156b predict a 1.10 $\rjup$ radius, which is discrepant with the Irwin et al. (2008) and Winn et al. (2009) determinations at roughly $1 \sigma$.  Our larger planet radius measurement of $1.095 \pm 0.020 ~\rjup$ ($1.0845 \pm 00.70 \rjup$) may lessen the need for substantial heavy-element enrichment, but as Winn et al. (2009) points out, the Fortney et al. (2007) models do not take into account tidal heating due to nonzero eccentricity, and the models are calculated for 4.5 Gyr, while the age of the system is estimated to be only 3 Gyr.  Each of these factors would likely increase the theoretical radius.

The observations presented in this study were scheduled as part of a major FGS program to detect stellar oscillations in HD 17156b.  The asteroseismology observations provided an independent constraint on the stellar density, which we found to be consistent with the transit determined value. The density constraint from asteroseismology has provided an extraordinary resource for refining parameter estimation via transit photometry.  The coexistence of asteroseismology observations and transit photometry anticipates the opportunities that \emph{Kepler} is expected to provide for a large number of transit hosting stars.

Support for Program GO-11945 was provided by NASA through a grant from
the Space Telescope Science Institute, which is operated by the
Association of Universitities for Research in Astronomy, Incorporated,
under NASA contract NAS5-26555.

\bibliographystyle{apj}

\end{document}